\documentclass[%
superscriptaddress,
notitlepage,
showpacs,
amsmath,amssymb,
aps,
longbibliography,
]{revtex4-1}

\usepackage{psfrag,graphicx,epsfig,color}
\usepackage{dcolumn}
\usepackage{bm}
\usepackage{natbib}
\usepackage{float}
\usepackage[usenames,dvipsnames,svgnames,table]{xcolor}
\usepackage{subfigure}
\usepackage{nicefrac}

\def\re    {{R_\lambda}}
\def\uu {{\mathbf{u}}}
\def\aa {{\mathbf{a}}}

\def\kk {{\mathbf{k}}}
\def\VV {{\mathbf{V}}}

\definecolor{mygreen}{rgb}{0,0.7,0.}

\begin{document}

\title{
Lagrangian acceleration in fully developed turbulence and its Eulerian decompositions
}

\author{Dhawal Buaria}
\email[]{dhawal.buaria@nyu.edu}
\affiliation{Tandon School of Engineering, New York University, New York, NY 11201, USA}
\affiliation{Max Planck Institute for Dynamics and Self-Organization, 37077 G\"ottingen, Germany}
\author{Katepalli R. Sreenivasan}
\affiliation{Tandon School of Engineering, New York University, New York, NY 11201, USA}

\date{\today}

\begin{abstract}

We study the properties of various Eulerian contributions to fluid particle 
acceleration by using well-resolved 
direct numerical simulations of isotropic turbulence, with the grid 
resolution as high as $12288^3$ and the Taylor-scale Reynolds number $\re$ 
in the range between 140 and 1300. 
The variance of convective acceleration, when normalized
by Kolmogorov scales, increases linearly with $R_\lambda$,
consistent with simple theoretical arguments, but very strongly differing from 
phenomenological predictions of Kolmogorov's hypothesis
as well as Eulerian multifractal models. The scaling of the local acceleration is also linear $R_\lambda$ to the leading order, but more complex in detail.
The strong cancellation between the local and convective 
acceleration -- faithful to the random sweeping
hypothesis -- results in the variance of the Lagrangian acceleration increasing only 
as $R_\lambda^{0.25}$,
as recently shown by Buaria \& Sreenivasan 
[Phys. Rev. Lett. 128, 234502 (2022)]. 
The acceleration variance is dominated by
irrotational pressure gradient contributions, whose variance
also follows an $R_\lambda^{0.25}$ scaling;
the solenoidal viscous contributions are relatively small and
follow a $R_\lambda^{0.13}$, consistent with Eulerian multifractal 
predictions. 

\end{abstract}

\maketitle


\section{Introduction}

In classical mechanics, the dynamics of particle motion is
characterized by the acceleration $\aa$, defined by
the rate of change of the particle
velocity $\uu$ in a Lagrangian frame. Given its fundamental role, 
the statistics of acceleration are of substantial interest in the 
study of turbulent flows
\cite{laPorta01, toschi:2009, stelzenmuller,buaria.rs, bec2006} and
also for stochastic modeling of transport phenomena
\cite{Sawford91, wyngaard, pope1994, wilson1996}.
Of particular interest is the scaling
of acceleration variance $\langle |\aa|^2 \rangle$ which, 
according to Kolmogorov's 1941 mean-field phenomenology \cite{K41a}, 
can be written as
$\langle |\aa|^2 \rangle = a_0 \langle \epsilon \rangle^{3/2}  \nu^{-1/2}$
\cite{MY.II},
where $\langle \epsilon \rangle$ is the mean dissipation rate,
$\nu$ is the kinematic viscosity and 
and $a_0$ is a universal constant. However, it is widely known
that due to small scale intermittency, $a_0$ is instead a variable that depends 
on the Reynolds number.
Following a few decades of investigations 
\cite{Yeung89, Vedula:99, gotoh99, Voth02, Sawford03, mordant2004, gylfason2004, yeung2006, Ishihara07},
recent data from direct numerical simulations (DNS) of isotropic turbulence
at high Reynolds numbers have established
that $a_0 \sim R_\lambda^{0.25}$ \cite{BS_PRL_2022}, where $\re$ is the
Taylor-scale Reynolds number. This result is at odds with
predictions from both Kolmogorov's theory and 
multifractal models \cite{borgas93, biferale2004}, 
and hence requires better understanding. 
In this Letter,
our aim is to further analyze this result, especially 
in terms of various underlying contributions to acceleration. 

We first note that, in fluid flows, an Eulerian viewpoint is often more
convenient, whereby the Lagrangian or material derivative is defined as
\begin{align}
    \aa = D\uu/Dt = \partial \uu /\partial t  +  \uu \cdot \nabla \uu  \ ,
\end{align}
where $\aa_L \equiv \partial \uu /\partial t$ is the local
component, capturing the unsteady rate of change at a fixed spatial position,
and $\aa_C \equiv \uu \cdot \nabla \uu$ 
is the convective component, capturing the rate of change 
due to spatial variations. That is, $\aa = \aa_L + \aa_C$. 
In addition, the dynamics of fluid motion in incompressible turbulent flows is
governed by the Navier-Stokes equations
\begin{align}
    \aa = -\nabla P +  \nu \nabla^2 \uu   \ , 
\end{align}
where $P$ is the kinematic pressure.
Since $\nabla \cdot \uu = 0$ from incompressibility,
the viscous term is solenoidal as well,
whereas the pressure gradient term is irrotational, i.e., its curl is zero. 
Thus, the acceleration
can also be written as $\aa = \aa_I + \aa_S$ 
(with suffixes $I$ and $S$ for irrotational and solenoidal, respectively), 
with $\aa_I \equiv -\nabla P$ and $\aa_S \equiv \nu \nabla^2 \uu$. 

In this Letter, we shall consider both methods of decomposition and study how
they relate to observed scaling of acceleration variance. Utilizing
data from the state-of-the-art DNS 
of isotropic turbulence, we show that the variance of convective acceleration 
varies as
$R_\lambda$, which follows from very simple theoretical arguments, but differs from 
multifractal predictions. The variance of local
component also varies as $R_\lambda$ to the leading order (with weaker
second order dependencies),
while always remaining slightly smaller than $\aa_C^2$. 
The Lagrangian acceleration results from strong cancellation between 
these two large quantities, varying as $R_\lambda^{0.25}$. 
We additionally explore how the properties $\aa_I$
and $\aa_S$ relate to local and convective
accelerations.

\section{Numerical approach and database}

The DNS data analyzed here
are obtained by solving the incompressible
Navier-Stokes equations,
corresponding to the canonical 
setup of forced stationary isotropic turbulence 
in a periodic domain \cite{Ishihara09, BPBY2019}.
Highly accurate Fourier pseudo-spectral methods
are utilized for spatial calculations, with aliasing errors controlled
using a combination of grid-shifting
and truncation \cite{Rogallo}. An explicit
second-order Runge-Kutta scheme is used for
time integration. 
The database for the present work is the same as
that of our recent study on acceleration \cite{BS_PRL_2022} 
and several other recent works
\cite{BS2020, BPB2020, BP2021, BP2022, BPB2022, BS2022}.
The grid resolution is as high as $12288^3$ and 
the Taylor-scale Reynolds number 
$\re$ lies in the range $140-1300$. Convergence with respect to 
small-scale resolution and statistical
sampling has been assessed in 
these previous studies.

As in \cite{BS_PRL_2022}, we have also calculated the relevant 
statistics using
Lagrangian fluid particle trajectories in the same
range of $\re$, albeit with lower small-scale resolution 
\cite{BSY.2015, BYS.2016, buaria.cpc}.
At the level of second order moments reported in this work,
the statistics are essentially identical from both Eulerian
and Lagrangian data. However, the Lagrangian particle
data are not suitable for studying higher order moments,
due both to the lack of resolution and accumulated numerical
errors resulting from interpolation of particle
velocities \cite{Yeung89}.

\section{Results}

\subsection{Theoretical analysis}

Before analyzing the DNS data, 
we present a brief theoretical analysis to obtain  
simplified relations between various Eulerian
components of acceleration. 
For instance, it is straightforward to prove that
in homogeneous turbulence, 
the correlation between an irrotational 
and a solenoidal vector is always zero
(see Appendix). From this property, 
it follows that
\begin{align}
  \label{eq:aias}
    \langle \aa_I \cdot \aa_S \rangle = 0  \\
    \langle \aa_I \cdot \aa_L \rangle = 0. 
    \label{eq:aial}
\end{align}
Additionally, using statistical stationarity, 
we can show (see Appendix) that
\begin{align}
    \langle \aa_S \cdot \aa_L \rangle = 0. 
    \label{eq:asal}
\end{align}
Since $\aa = \aa_I + \aa_S$, it also follows from 
Eqs.~\eqref{eq:aial} and \eqref{eq:asal} that
\begin{align}
    \langle \aa \cdot \aa_L \rangle = 0 \ , 
    \label{eq:atal}
\end{align}
i.e., the Lagrangian acceleration $D\uu/Dt$ is uncorrelated
to the Eulerian acceleration $\partial u/\partial t$. 
This property directly yields the following result:
\begin{align}
    \langle \aa_L \cdot \aa_C \rangle = - \langle |\aa_L|^2 \rangle.
\end{align}
These relations allow us to write the
acceleration variance as
\begin{align}
\label{eq:a2is}
\langle |\aa|^2 \rangle &= \langle |\aa_I|^2 \rangle + \langle |\aa_S|^2 \rangle \\
\langle |\aa|^2 \rangle &= \langle |\aa_C|^2 \rangle - \langle |\aa_L|^2 \rangle.
\label{eq:a2lc}
\end{align}
Thus, while the acceleration variance is given by 
the sum of variances of the pressure gradient and 
viscous terms,
it is also obtained via a direct cancellation of convective 
and local components.
We will now explore how the scaling of all these Eulerian contributions affect the scaling of acceleration variance itself.

\subsection{Properties of $\aa_I$ and $\aa_S$}

\begin{figure}
\begin{center}
\includegraphics[width=0.41\linewidth]{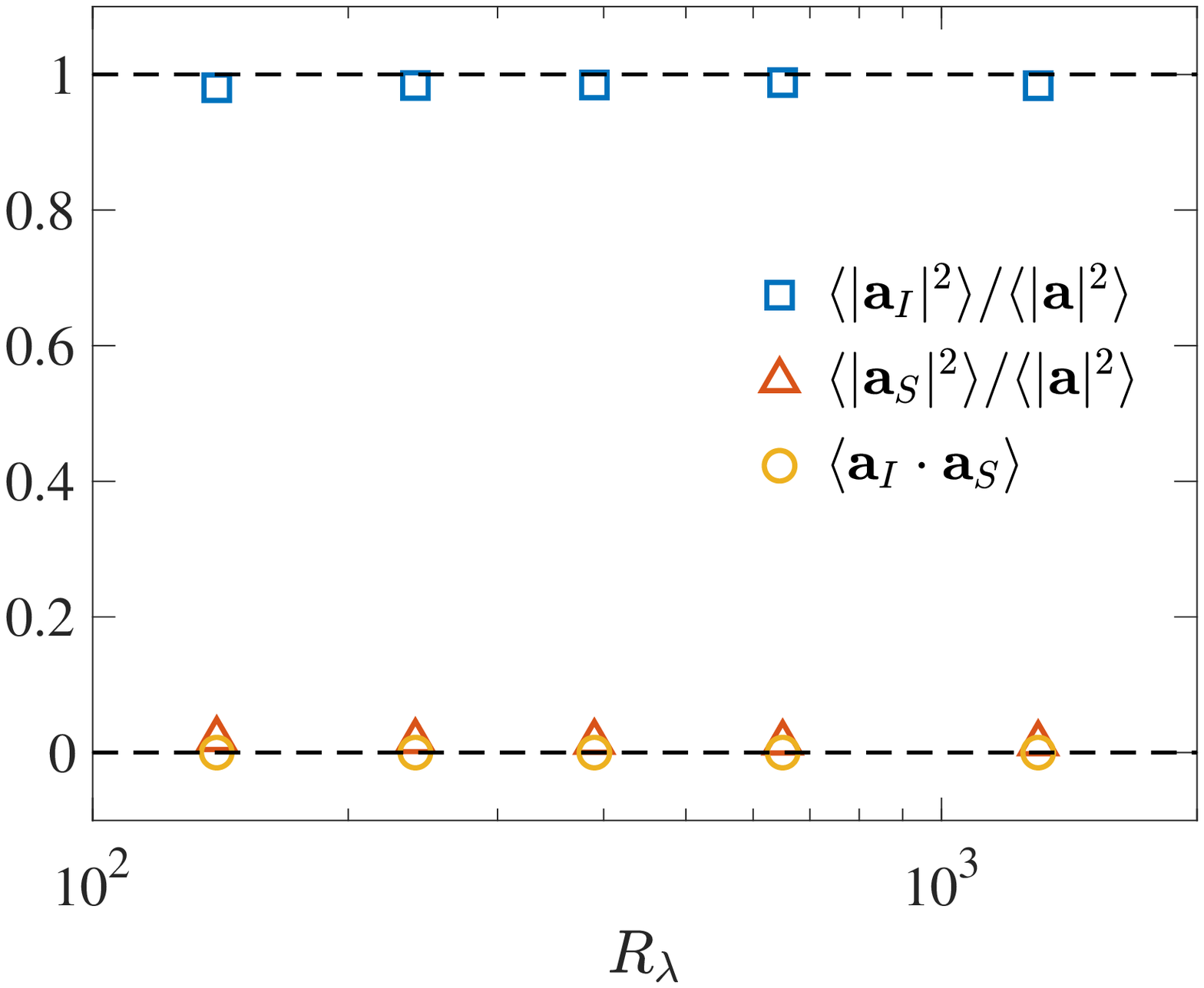} \ \ \ \
\includegraphics[width=0.43\linewidth]{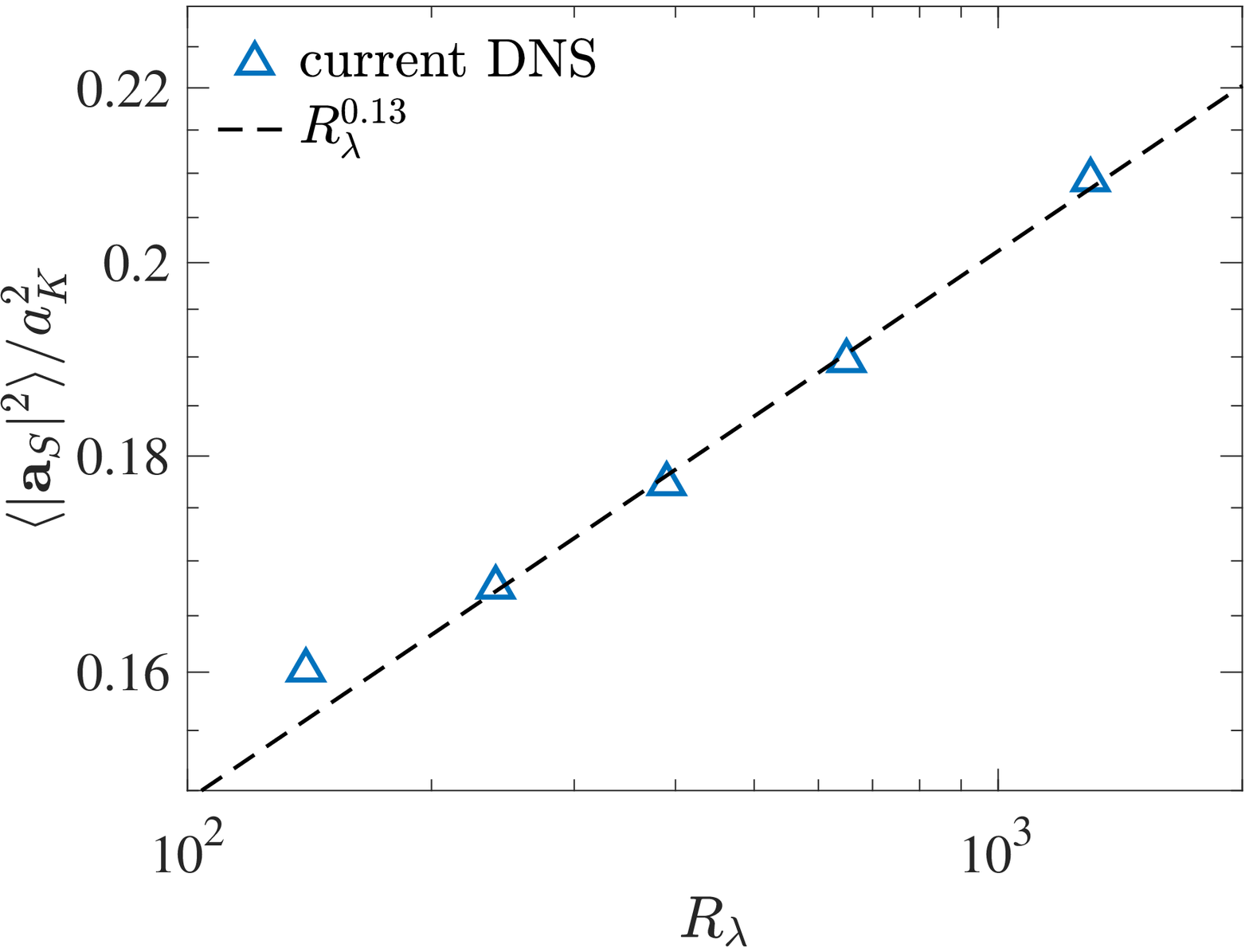} 
\caption{
(a) Fractional contributions 
of the irrotational pressure gradient and solenoidal viscous 
terms to the acceleration variance, as well as their mutual correlation,
as functions of $\re$.
(b) Variance of the solenoidal viscous acceleration
normalized by the Kolmogorov scales, as a function of $\re$.
}
\label{fig:aias}
\end{center}
\end{figure}

It is well known that acceleration variance is 
dominated by the irrotational pressure gradient contribution
and the corresponding viscous contribution is negligible,
i.e., $|\aa_I| \gg |\aa_S|$ \cite{Vedula:99, tsinober01}.
We first reaffirm this result in Fig.~\ref{fig:aias}a which
shows the fractional contributions
of $\aa_I$ and $\aa_S$, and also 
the correlation $\langle \aa_I \cdot \aa_S \rangle$, which is zero as expected.
It is evident that  $\langle |\aa|^2 \rangle \approx \langle |\aa_I|^2 \rangle$.
We can readily show that
\begin{align}
\langle \aa \cdot \aa_I \rangle = \langle \aa_C \cdot \aa_I \rangle = \langle |\aa_I|^2 \rangle \ , 
\ \ \ \ 
\langle \aa \cdot \aa_S \rangle = \langle \aa_C \cdot \aa_S \rangle = \langle |\aa_S|^2 \rangle \ ,
\end{align}
which demonstrate that both the pressure gradient and viscous contributions
predominantly arise from the convective component. This is not surprising
since we saw earlier that  
$\langle \aa_I \cdot \aa_L \rangle$ and 
$\langle \aa_S \cdot \aa_L \rangle$ are both zero.
We shall further elaborate this point in the next subsection. 

It is worth noting that, while the contribution
from the viscous term $\aa_S$ is negligible in comparison to $\aa_I$, 
it is nevertheless finite and intimately connected
to the fundamental dynamics of turbulence. 
In particular, its variance can be written as \cite{MY.II, Vedula:99} 
\begin{align}
    \langle |\aa_S|^2 \rangle / a_K^2 =  -\frac{35}{2} \frac{\mathcal{S}}{(15)^{3/2}} \ , 
    \label{eq:as2}
\end{align}
where $a_K = \langle \epsilon \rangle^{3/4} \nu^{-1/4}$
and $\mathcal{S}$ is the skewness of longitudinal velocity 
gradients, which is always negative in turbulence,
characterizing the energy cascade from large to
small scales \cite{Batchelor,Frisch95}. The skewness can also
be related to vortex stretching \cite{Betchov56}
and is known to weakly increase in magnitude as $R_\lambda^{0.13}$ 
\cite{BBP2020}. This scaling matches the prediction
from extending Eulerian multifractals to Lagrangian variables \cite{Sreeni88,Frisch95,borgas93}. Figure~\ref{fig:aias}b shows a satisfactory agreement with this result 
(except at the lowest Reynolds number).

However, our recent work \cite{BS_PRL_2022} computed the acceleration variance $\langle |\aa_I|^2 \rangle$ and found it to vary as $R_\lambda^{0.25}$. We expressed the acceleration  analytically in terms of the fourth order velocity structure functions \cite{hill1995, hill2002}, and showed that
\begin{align}
\langle |\aa|^2 \rangle /a_K^2 \approx \langle |\aa_I|^2 \rangle /a_K^2 \sim R_\lambda^{0.25} \ .
\label{eq:ai2}
\end{align}
The data from various sources, including our own DNS, 
show excellent agreement with this prediction 
(see \cite{BS_PRL_2022}; we also reaffirm it below). It then follows that an 
extension of Eulerian multifractals to explain 
intermittency of Lagrangian 
quantities is fraught with major uncertainties.


\subsection{Properties of $\aa_L$ and $\aa_C$}

From Eq.~\eqref{eq:a2lc}, acceleration variance
results from direct cancellation between the variances of
$\aa_C$ and $\aa_L$. 
This cancellation is consistent with the random sweeping hypothesis 
proposed by Kraichnan \cite{kraichnan64} -- see also Tennekes \cite{tennekes75} -- 
which states that 
the small scales of turbulence 
are swept past an Eulerian observer on a much shorter time scale
than the time scale governing their dynamical evolution. 
The nominal validity of this hypothesis is also implicitly reflected in the fact
that $\aa = D\uu/Dt$ and $\aa_L = \partial \uu/\partial t $ 
are uncorrelated (see Eq.~\eqref{eq:atal}).

The convective acceleration $\aa_C = \uu \cdot \nabla \uu$
essentially represents a correlation between
the velocity and its gradients.
Given the general understanding that the former characterizes the large scales
and the latter the small scales, we can assume that the two
are essentially uncorrelated (provided $R_\lambda$ is sufficiently high). 
Thus, simple scaling arguments
suggest that $|\aa_C| \sim u'/ \tau_K$ where
$u' $ is the root-mean-square (rms) velocity 
and $\tau_K$ is the Kolmogorov time scale, characterizing
the rms of velocity gradients.
We then have
\begin{align}
    \langle |\aa_C|^2\rangle /a_K^2 = c \ R_\lambda \ , 
    \label{eq:ac2}
\end{align}
where $c$ is some proportionality constant and we have utilized the classical
estimate $u'/u_K \sim R_\lambda^{1/2}$
\cite{Frisch95} ($u_K$ being the Kolmogorov velocity scale).
To the first order, it can be also expected that 
$\langle |\aa_L|^2\rangle$ also follows a similar scaling. 
This can be inferred indirectly by noting that
\begin{align}
\langle |\aa_L|^2\rangle &= \langle |\aa_C|^2\rangle - 
    \langle |\aa|^2\rangle   \\ 
     &= \langle |\aa_C|^2\rangle - 
    \langle |\aa_I|^2\rangle - \langle |\aa_S|^2\rangle. 
    \label{eq:al2}
\end{align}
where observations show that the scaling of $|\aa_C|^2$ 
dominates over other two components.

\begin{figure}
\begin{center}
\includegraphics[width=0.43\linewidth]{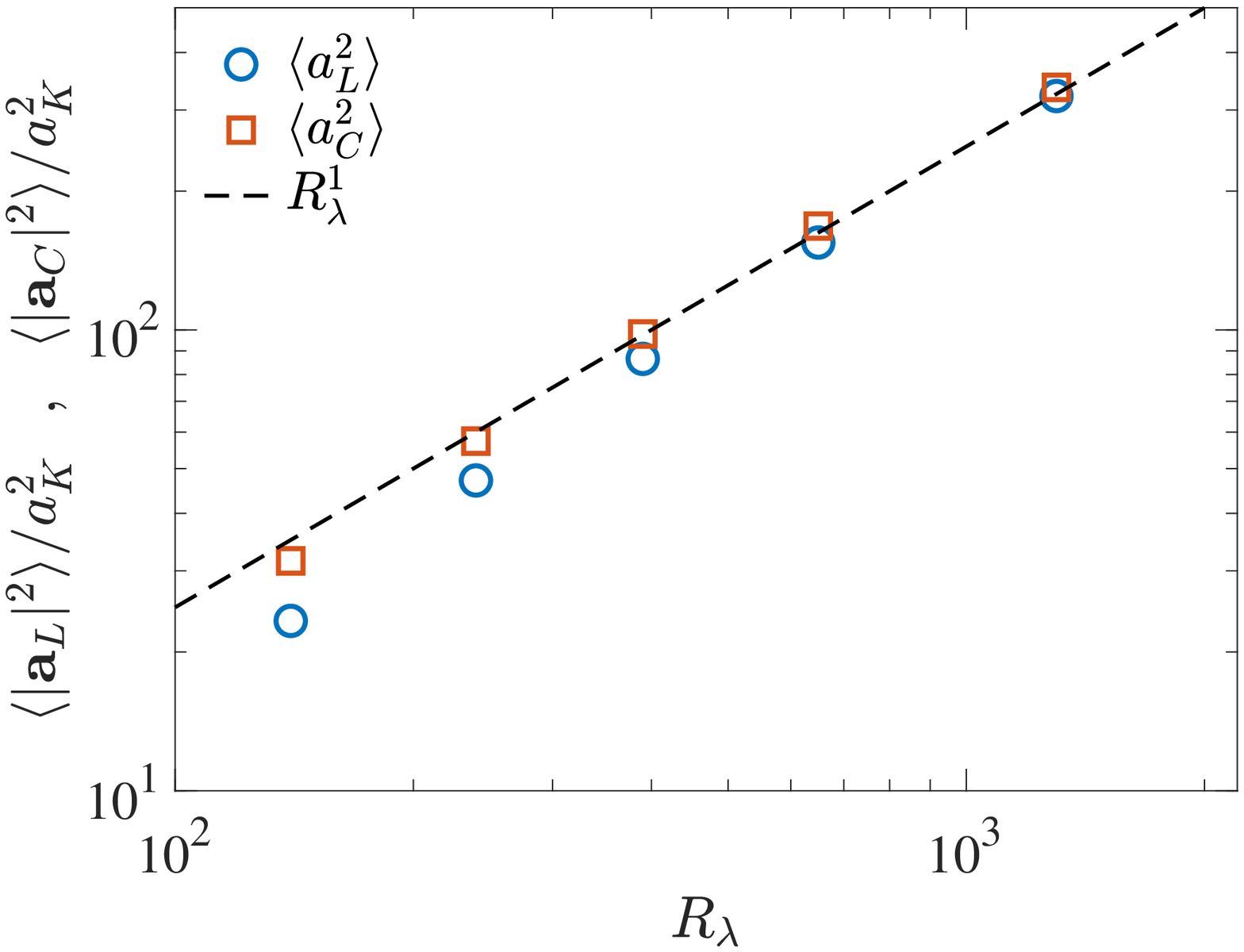} \\
\vspace{0.2cm}
\includegraphics[width=0.43\linewidth]{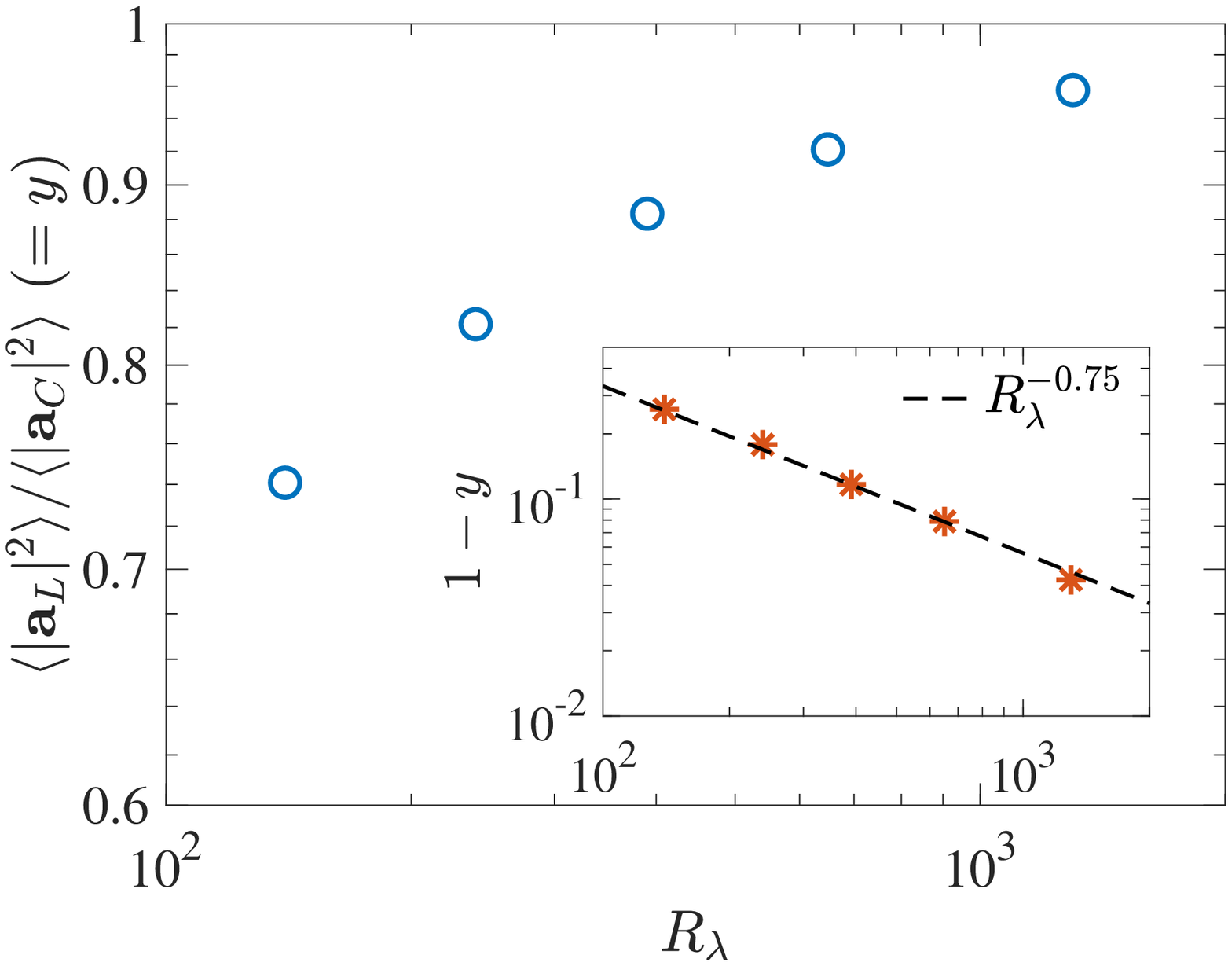} \ \ \ \ \
\includegraphics[width=0.43\linewidth]{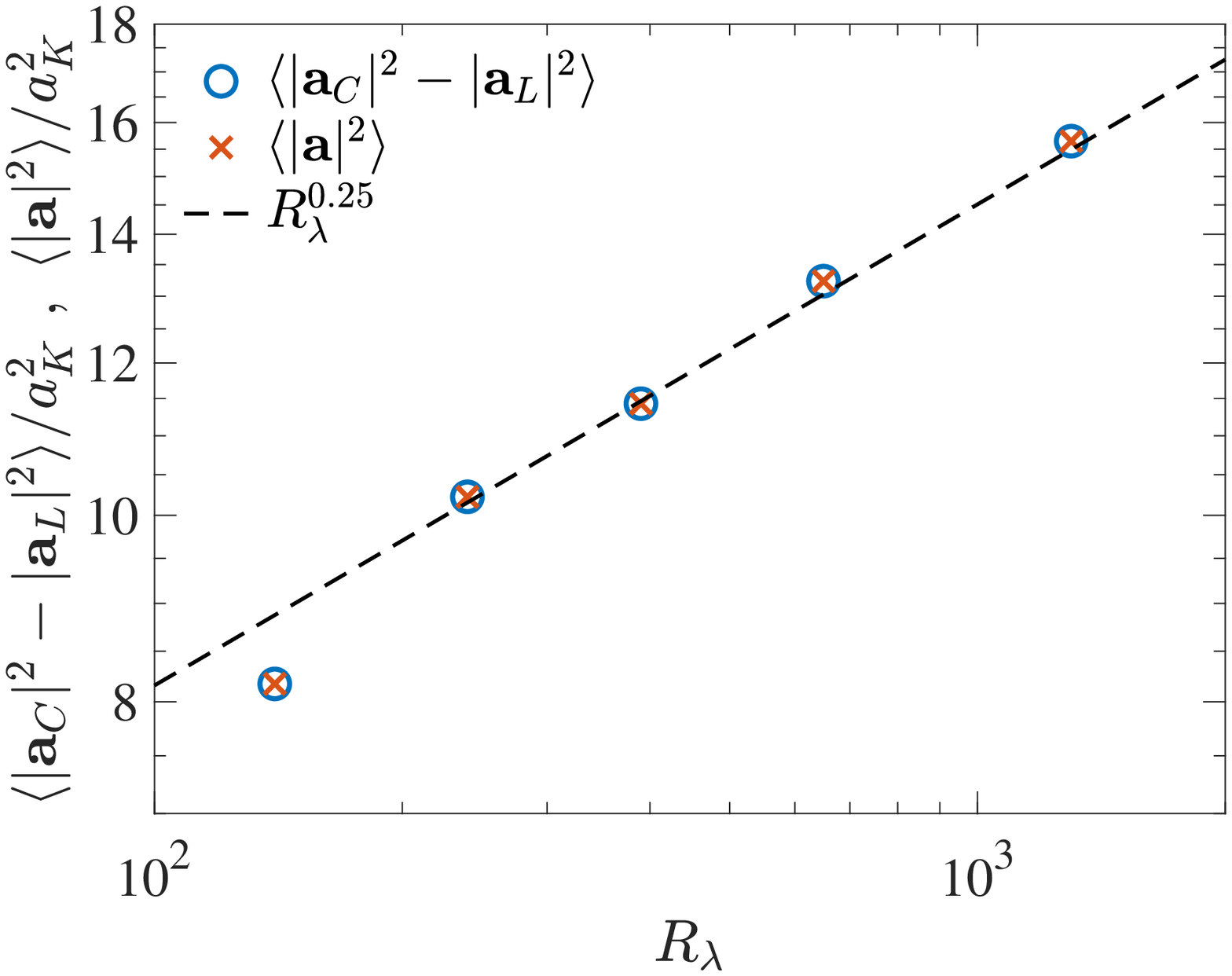} 
\caption{
(a) Variances of local and convective acceleration,
normalized by Kolmogorov scales, as a function of $\re$.
(b) Ratio of variance of local and convective acceleration;
inset shows deficit of the ratio from unity
(c) Difference between the convective and local 
acceleration as a function of $\re$, compared
directly with acceleration variance.
}
\label{fig:alac}
\end{center}
\end{figure}

Figure~\ref{fig:alac}a shows the variances of local
and convective acceleration. It can be immediately seen that 
$|\aa_C|^2/a_K^2$ follows
a simple linear scaling in $R_\lambda$, as anticipated in Eq.~\eqref{eq:ac2}.
On the other hand, $|\aa_L|^2/a_K^2$ approaches this scaling
as $\re$ increases, but noticeably deviates at low $\re$.
Using the results in Eqs.~\eqref{eq:as2}-\eqref{eq:al2},
the precise scaling of $\langle |\aa_L|^2 \rangle$
can be quantified as
\begin{align}
\langle |\aa_L|^2\rangle /a_K^2 = 
c R_\lambda   - c_1 R_\lambda^{0.25} - c_2 R_\lambda^{0.13} \ , 
\label{eq:al2_re}
\end{align}
where $c_1$ and $c_2$ are proportionality constants. 
Evidently, the deviations at lower $\re$ can be understood
in terms of these additional terms.
It also follows from here that the ratio $\langle |\aa_L|^2\rangle / \langle |\aa_C|^2\rangle$ 
has the form $1 -(c_1/c)R_\lambda^{-0.75} - (c_2/c) R_\lambda ^{-0.87}$. Asymptotically, when normalized by Kolmogorov variable, $\langle |\aa|^2 \rangle = \langle |\aa_C|^2 \rangle - \langle |\aa_L|^2 \rangle \approx c_1 R_\lambda^{-0.25}$.

To verify the behaviors of $\aa_L$ and $\aa_C$
we plot in 
Fig.~\ref{fig:alac}b the ratio 
$y = \langle |\aa_L|^2\rangle / \langle |\aa_C|^2\rangle$ 
as a function of $\re$. The inset
shows $1-y$, {which is in excellent agreement}
with a power law $R_\lambda^{-0.75}$. The ratio 
steadily approaches unity, which demonstrates that, asymptotically, 
only the $R_\lambda^{0.25}$ term contributes to Lagrangian acceleration. 
This is also confirmed by Fig.~\ref{fig:alac}c,
which shows
$\langle |\aa_C|^2\rangle  - \langle |\aa_L|^2\rangle$
normalized by Kolmogorov scales
with the acceleration variance data from \cite{BS_PRL_2022} -- 
both sets of points are indistinguishable and
in excellent agreement with  $R_\lambda^{0.25}$ scaling.


\subsection{Further analyzing the role of $\aa_C$}

The near cancellation between $\aa_L$ and $\aa_C$
can be further analyzed by noticing that 
$\aa_L$ is solenoidal, whereas $\aa_C$ is not; thus, they 
can never completely cancel each other.
Further, since 
$\aa_I$ is irrotational and $\aa_S$ is solenoidal, 
we can write \cite{tsinober01}
\begin{align}
    \aa_{C_I} &= \aa_I \ ,  \\ 
    \aa_L + \aa_{C_S} &= \aa_S \ ,
\end{align}
where we have decomposed $\aa_C$ into
irrotational and solenoidal components, i.e., 
$\aa_C = \aa_{C_I} + \aa_{C_S}$.
Such a decomposition can readily be implemented
in Fourier space using the Helmholtz decomposition, i.e.,
for a vector $\mathbf{V}$ with Fourier
coefficient $\hat{\mathbf{V}}$, 
the Fourier coefficients of irrotational
and solenoidal parts are, respectively, given as
\begin{align}
    \hat{\VV}_I (\kk) = (\kk \cdot \hat{\VV}) \kk / k^2 \ ,
    \ \ \ \   
    \hat{\VV}_S (\kk) = \hat{\VV} - \hat{\VV}_I \ ,
\end{align}
where $\kk$ is the wave-vector and $k = |\kk|$. 
Note that irrotationality is 
imposed in Fourier space by the condition
$\kk \times \hat{\VV}_I = \mathbf{0}$,
and solenoidality by
$\kk \cdot \VV_S = 0$
(both of which can be easily verified). 

\begin{figure}
\begin{center}
\includegraphics[width=0.43\linewidth]{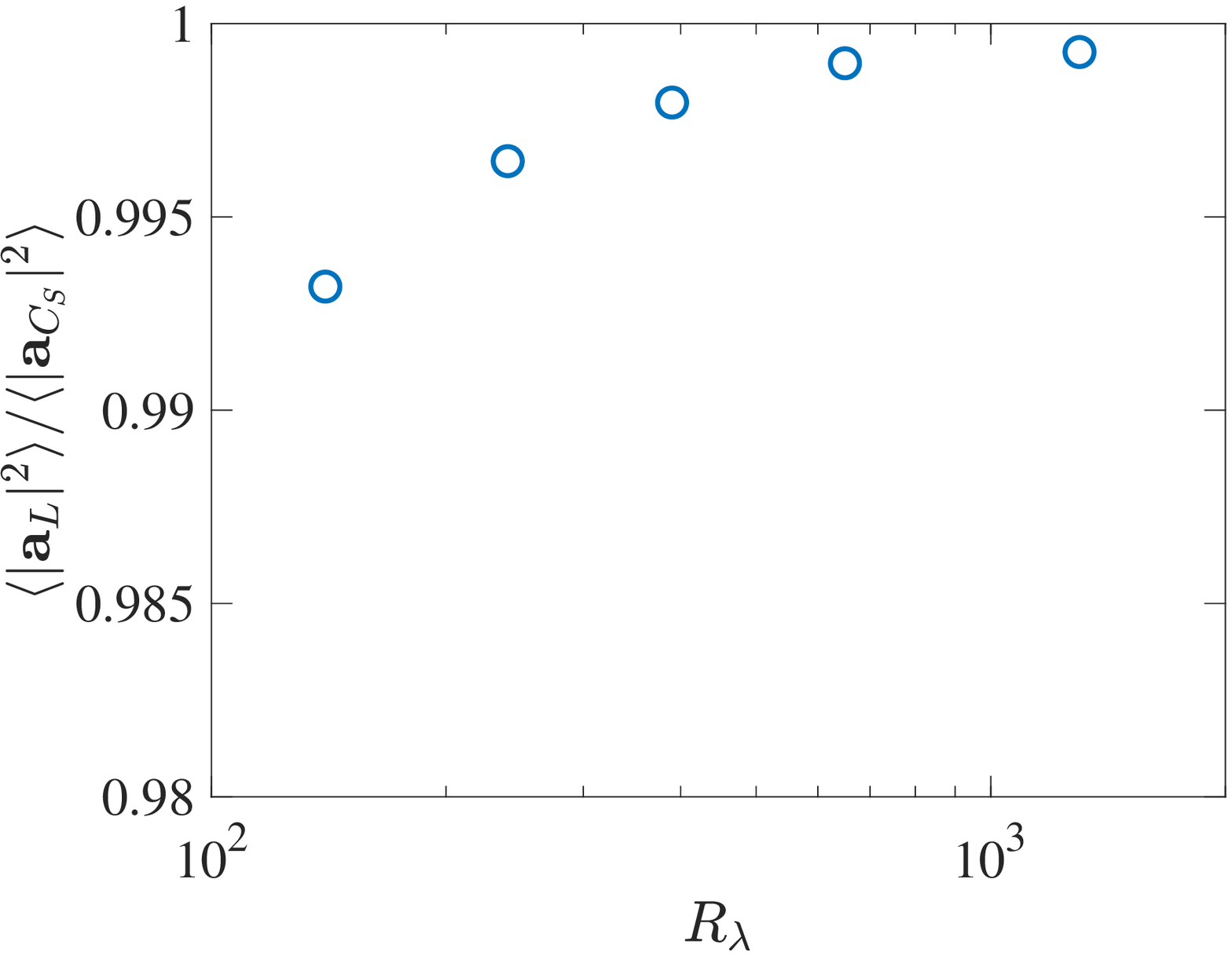} \ \ \ \
\includegraphics[width=0.43\linewidth]{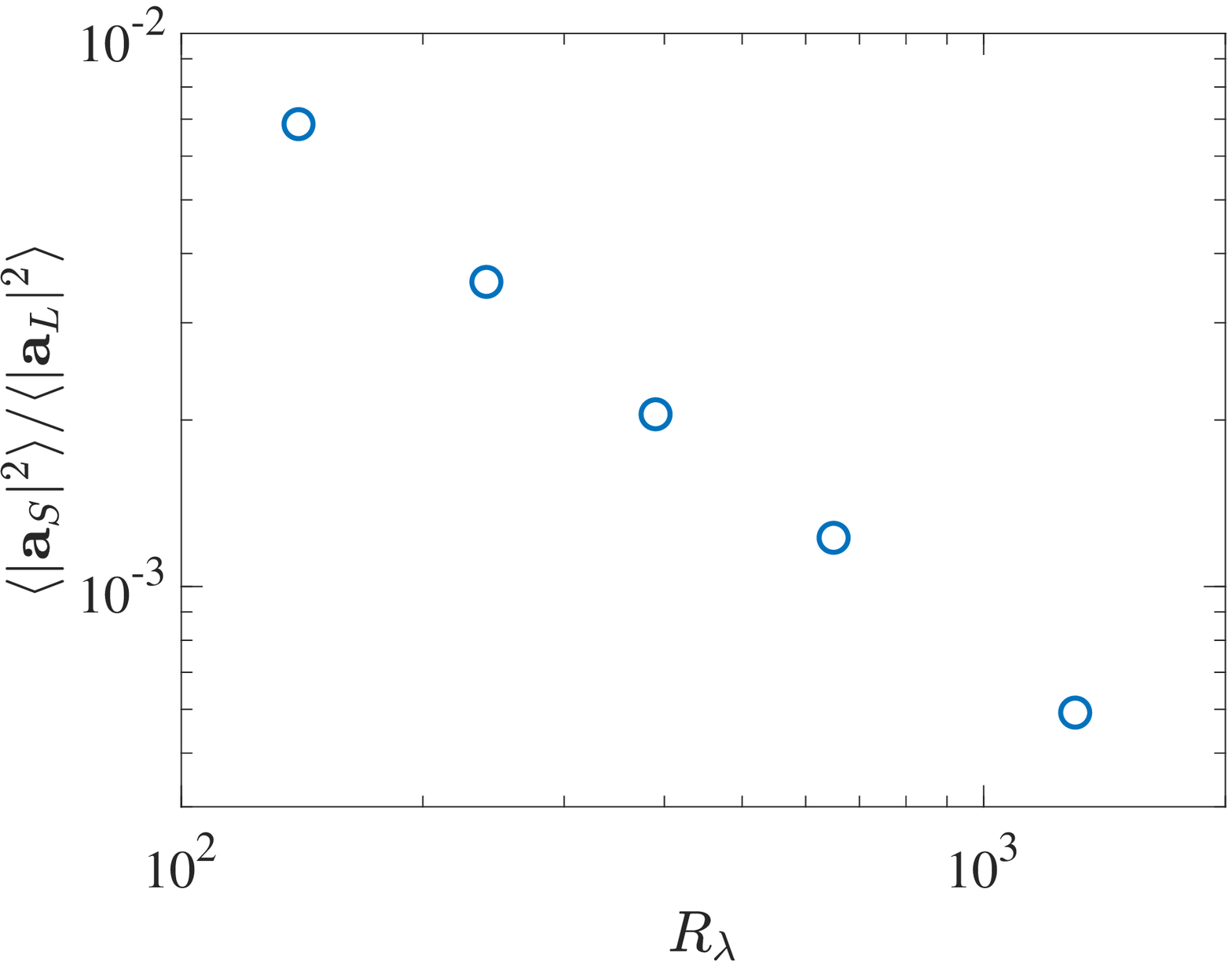} 
\caption{
(a) Ratio of variance of local acceleration to 
that of solenoidal component of convective acceleration,
as a function of $\re$.
(b) Ratio of variance of solenoidal viscous acceleration
to that of local acceleration. 
}
\label{fig:acs}
\end{center}
\end{figure}

From the above decomposition, 
it trivially follows that
$ |\aa_{C_I}|^2 = |\aa_I|^2 $ 
and $ \langle |\aa_{C}|^2 \rangle  = \langle |\aa_{C_I}|^2 \rangle  + 
\langle |\aa_{C_S}|^2 \rangle$  
and we can also show that
\begin{align}
   \langle \aa_L \cdot \aa_{C_S} \rangle &= 
   - \langle |\aa_L|^2 \rangle  \ ,  \\
   \langle |\aa_S|^2 \rangle &= \langle |\aa_{C_S}|^2 \rangle 
   - \langle |\aa_{L}|^2 \rangle \ . 
\end{align}
Thus, the very small solenoidal component $\aa_S$ 
results from near perfect cancellation between
$\aa_{C_S}$ and $\aa_L$. We quantify this 
in Fig.~\ref{fig:acs}. Panel a shows that 
the ratio $\langle |a_L|^2 \rangle / \langle |a_{C_S}|^2 \rangle$ steadily approaches 
unity as $\re$ increases
(though it cannot be strictly unity since $|\aa_S|$ is always
finite). 
In panel b, the ratio 
$\langle |a_S|^2 \rangle / \langle |a_{L}|^2 \rangle$ 
is shown, which steadily decreases with $\re$ as expected.

In summary,
based on the present analysis, we can essentially write
$|\aa_C| \gtrsim |\aa_{C_S}| \gtrsim |\aa_L| \gg |\aa| \approx |\aa_I| = |\aa_{C_I}| \gg |\aa_S|$. Perhaps surprisingly, $\langle |\aa|^2 \rangle \approx \langle |\aa_I|^2 \rangle \sim R_\lambda^{0.25}$, while variances of the components $\aa_C$ and $\aa_L$ have far stronger dependencies on $R_\lambda$; for example, $\langle |\aa_C|^2 \rangle$  essentially
scales as $R_\lambda$.

\section{conclusions} 
The most interesting result of the analysis is that 
the local and convection terms of acceleration are anti-correlated and both of them depart from Kolmogorov's paradigm very strongly (also from the multifractal formalism). In particular, the 
variance of both essentially scale linearly with $R_\lambda$. The two terms are, however, strongly anti-correlated. Thus, the difference between the two, which specifies the Lagrangian acceleration, increases as $R_\lambda^{0.25}$ \cite{BS_PRL_2022}. This result, which is an indication that the two terms are separately much more intermittent than their algebraic (vector) sum, is at odds with Kolmogorov's and multifractal formalisms, but not nearly as much as their sum. The scaling 
$\langle |\aa_C|^2 \rangle \sim R_\lambda$ comes from the assumption that $\bf u$ and  $\nabla \bf u$ are uncorrelated, so essentially $\langle |\aa_C|^2 \rangle$ follows the same scaling as 
$\langle |\uu|^2 \rangle$. The interpretation is that the small scales are simply swept by the large scale velocity without getting affected, which is consistent with the random sweeping hypothesis.

\appendix
\section{Vanishing correlations between various Eulerian contributions to 
acceleration}

Let us assume that vector $\mathbf{A}$ is irrotational and
vector $\mathbf{B}$ is solenoidal; this implies 
$\nabla \times A = \mathbf{0}$ and $\nabla \cdot B=0$ (or $\partial B_i/\partial x_i = 0$).
For the former, we can write $A_i = \partial \phi /\partial x_i$, where $\phi$ is some
scalar quantity. Thus, the correlation can be simplified as:
\begin{align}
    \langle \mathbf{A} \cdot \mathbf{B} \rangle 
    &= \left\langle  \frac{\partial \phi}{\partial x_i} B_i \right\rangle \\ 
    &= \left\langle  \frac{ \partial (\phi B_i)}{\partial x_i} -  \phi \frac{\partial B_i}{\partial x_i} \right\rangle \\
    &=  \frac{ \partial \langle \phi B_i \rangle}{\partial x_i} -  \left\langle  \phi 
    \frac{\partial B_i}{\partial x_i} \right\rangle \\
    & = 0,
\end{align}
where the first term is zero from statistical homogeneity and the second term is zero
since $\partial B_i/\partial x_i$.

Thus, for the components of acceleration, we can write
\begin{align}
    \langle \aa_I \cdot \aa_S \rangle = 0 \\ 
    \langle \aa_I \cdot \aa_L \rangle = 0. 
\end{align}

For the correlation between $\aa_S$ and $\aa_L$, the following
steps have to be considered:
\begin{align}
    \langle \aa_S \cdot \aa_L \rangle 
    &=  \left\langle \nu \frac{\partial^2 u_i}{\partial x_k \partial x_k}  \cdot
        \frac{ \partial u_i}{\partial t} \right \rangle \\ 
    &=  \nu \left\langle  \frac{\partial}{\partial x_k} 
     \left( \frac{\partial u_i}{\partial x_k} \frac{\partial u_i }{\partial t} \right)
    - \frac{\partial u_i}{\partial x_k} \cdot \frac{\partial}{\partial t} \left( \frac{\partial u_i}{\partial x_k} \right) \right\rangle \\
    &=  \nu \left\langle  \frac{\partial}{\partial x_k} 
     \left( \frac{\partial u_i}{\partial x_k} \frac{\partial u_i }{\partial t} \right) \right\rangle
    -  \frac{\nu}{2} \left\langle  \frac{\partial}{\partial t} \left( \frac{\partial u_i}{\partial x_k} \right)^2 \right\rangle \\
    &=0.
\end{align}
The last step follows from the fact that the first term is zero from statistical homogeneity, whereas the second term 
is zero from statistical stationarity. Since $\aa = \aa_I + \aa_S$, it also follows that 
\begin{align}
   \langle \aa \cdot \aa_L \rangle = 0.  
\end{align}


%

\end{document}